\documentclass[letterpaper]{article}
\usepackage{flairs}
\usepackage{times}
\usepackage{helvet}
\usepackage{courier}
\usepackage{adjustbox}
\usepackage{amsmath,amssymb,amsfonts}
\usepackage{algorithmic}
\usepackage{graphicx}
\usepackage{textcomp}
\usepackage{xcolor}
\usepackage{url}
\usepackage{xurl}
\begin{document}
%
\title{Deep Learning Approach for Early Stage Lung Cancer Detection}
\author{Saleh Abunajm, Nelly Elsayed, Zag ElSayed, Murat Ozer\\
School of Information Technology\\
University of Cincinnati\\
Ohio, United States\\
}
\maketitle
\begin{abstract}
\begin{quote}
Lung cancer is the leading cause of death among different types of cancers. Every year, the lives lost due to lung cancer exceed those lost to pancreatic, breast, and prostate cancer combined. The survival rate for lung cancer patients is very low compared to other cancer patients due to late diagnostics. Thus, early lung cancer diagnostics is crucial for patients to receive early treatments, increasing the survival rate or even becoming cancer-free. This paper proposed a deep-learning model for early lung cancer prediction and diagnosis from Computed Tomography (CT) scans. The proposed mode achieves high accuracy. In addition, it can be a beneficial tool to support radiologists' decisions in predicting and detecting lung cancer and its stage. 
\end{quote}
\end{abstract}

\section{Introduction}
\noindent  Cancer is a disease that can affect any part of the body and cause uncontrolled, abnormal cell growth. The characteristics of cancer cells are different from normal cells, such as uncontrollable growth, invading and spreading to nearby tissues, and they are immortal. 

Lung cancer is the most common cancer worldwide~\cite{wcrf}, and it is the third most common cancer in the UK and the United States. The leading killer of cancer in men and women is lung cancer. Almost 25\% of all cancer deaths are caused by lung cancer. It is most common in older people. Most people diagnosed with lung cancer are 65 or older. A small number of individuals diagnosed with lung cancer are younger than 45~\cite{acs}. Based on the death numbers that statistics have shown, lung cancer is responsible for almost 25\% of deaths caused by cancer. There are five stages of cancer~\cite{woodard2016lung,hammerschmidt2009lung}:
\begin{itemize}
	\item \textbf{Stage 0:} This is the earliest Stage of cancer, and the cancer cell still did not spread.
	\item \textbf{Stage I:} Cancer cells are small and spread into local areas but do not spread to nearby lymph nodes or other body parts.
	\item \textbf{Stage II:} Cancer cells have spread into nearby lymph nodes or tissues in a local area. In this Stage, cancer cells are more significant than in Stage I.
	\item \textbf{Stage III:} Cancer cells spread to a set of lymph nodes, and the tumor growth exceeds a specific size.
	\item \textbf{Stage IV:} In this Stage, the cancer is considered metastatic; cancer cells have invaded other organs and spread to other parts of the patient’s body.
	The acknowledgment is hidden to maintain the anonymity of the authors and their affiliations. The full acknowledgment will be added to the final version.
\end{itemize}

 It is critical to continue the research to help in the early detection of lung cancer. The American Cancer Society’s estimates for lung cancer in the United States for 2021 are~\cite{acs}:
\begin{itemize}
	\item About 235,760 new cases of lung cancer (119,100 in men and 116,660 in women).
	
	\item About 131,880 deaths from lung cancer (69,410 in men and 62,470 in women).
\end{itemize}

\begin{table*}[t]
	\begin{center}
			\resizebox{\textwidth}{!}{	
			\begin{tabular}{|l|c|l|l|l|l|l|}
				\hline
				\textbf{Author} & \textbf{Year} & \textbf{Model} & \textbf{Dataset} &\textbf{Imaging Type}& \textbf{Remarks}& \textbf{Limitations} \\ 
				\hline 
				\cite{teramoto2016automated} & 2016 &CNN & 104 Japanese men and women & CT and PET images & False positive was reduced & A small number of patients \\
				 &   & & 104 Japanese men and women & CT and PET images & compared to their previous study & were used in the study. \\
				\hline
				\cite{van2015off} & 2015 &  OverFeat CNN& LIDC dataset 865 CT scans  & CT scan & Combining OverFeat CNN with CAD indicated  & Using OverFeat CNN achieved  \\
				  &  &   &  &   & remarkable improvement in detection performance &  unfavorable results compared to CAD  \\				
				\hline
				
				\cite{ali2018lung} & 2018 & CNN Reinforcement Learning& LIDC/IDRI database (LUNA)  & CT scan &The model does not  & Small number of CT scans  \\	
				
			 &   &  &  challenge, 888 CT scans  & CT scan &require preprocessed data & Small number of CT scans  \\	
				\hline
				
				\cite{shen2019interpretable} & 2019 &  Hierarchical semantic convolutional& LIDC-IDRI (1010 patients)
				& CT scan & HSCNN accomplished better& Semantic labels did not\\
				
			  &  &    neural network (HSCNN)& LIDC-IDRI (1010 patients)
				& CT scan &performance than a 3D CNN &contain nodule size, location,  \\
				
				&  &  & 
				&  & &or margin spiculation   \\
				
				\hline
				
				\cite{zhang2019lung} & 2019 &VGG16, VGG19, ResNet50,& LIDC-IDRI & CT scan &DenseNet121 and Xception
				have 
				&Small sample size\\ 
				
				& &DenseNet121, MobileNet, Xception,&(1018 CT scans/1010 patients) &  &better results detecting lung nodule
				&used in the study \\ 
				
				& &NASNetMobile, and NASNetLarge& &  &DenseNet121 and Xception
				
				&\\ 			
				\hline				
				\cite{song2017using} &  2017 & CNN,DNN, and SAE& LIDC-IDRI (4581 images) & CT scan &CNN has a better performance&Due to the dataset’s limitations, \\
				
				 &  & &  &  &compared to DNN and SAE&the CNN architecture has small layers \\
		
				\hline
				\cite{sahu2018lightweight} & 2018 &  Multi-Section CNN& LIDC-IDRI dataset (649 patients) & CT scan & It takes one hour to train the MobileNet based model &Small cohort used  in the study  \\
				\hline
			\end{tabular}
			}
		\caption  {A comparison between the state-of-the-art CNN based models for detection lung cancer.}
		\label{compare_table}
	\end{center}
\end{table*}
Medical providers use different types of cancer scans to help diagnose and plan treatment. The most common types are magnetic resonance imaging (MRI) scans, which take detailed images of the patient’s target area in the body; computed tomography (CT) scans, also known as CAT Scans, which create 3D images of the patient’s target area in the body from different angles, and positron emission tomography (PET/CT), which uses a tracer gets injected into the patient’s body. It makes the cancerous cells appear brighter than the non-cancerous cells.

The survival rate in each stage is different, and the earlier medical providers diagnose cancer, the higher the survival rate. While we keep in mind that doctors do not have a guaranteed treatment to cure cancer, diagnosing cancer at an early stage gives medical providers more time for their treatment plans. The treatment for a stage IV patient is particularly challenging. Therefore, computing tools can be used to support medical providers in early lung cancer diagnostics.

This paper proposed a Deep Convolutional Neural Network (CNN) based model for the early prediction and detection of lung cancer and its stages. The proposed model output testing is to be comparable to real-life cases diagnosed by expert radiologists. Thus, it can be an effective tool to assist medical providers in diagnostic decision-making. 

\section{Related Work}

Medical imaging refers to several different technologies that are used to view the human body to diagnose, monitor, or treat medical conditions. There are several types of medical imaging, such as magnetic resonance imaging (MRI), CT scans, also known as CAT Scans, positron emission tomography (PET/CT), X-ray, arthrogram, ultrasound, and myelogram. For lung cancer, a CT scan is the recommended type of medical imaging. CT scans point out whether abnormal growth exists in the lung. This abnormal growth could be cancerous (malignant) or non-cancerous (benign). Therefore, radiologists review CT scans to determine if cancer has developed. Detecting lung cancer is extremely important. However, this is challenging for radiologists as the lung nodules can be missed or even some false-positive diagnoses. Lung cancer classification is crucial. In deep learning, nodules classification could be based on nodules’ location, size, number, consistency, and other factors.

In 2022, Wenfa et al.~\cite{jiang2022application} proposed a deep learning model to verify the prediction accuracy of lung cancer using CT images. In their study, they used two types of images formats, ‘.DICOM’ and ‘.MHD’ formats. One of the main highlights of their study is false positive reduction, and they used U-Net and 3D CNN which achieved high accuracy in false-positive nodules screening. In 2020, Tasnim Ahmad et al.~\cite{ahmed2020lung} used 3D CNN classifier. The proposed CNN model has two conventional layers, two max-pooling layers, and a fully connected layer. LUNA data set was used with one hundred (100) patients divided between training and testing 80\%, 20\% respectively. The accuracy rate was 80\% on 400 images tested. In 2020, A hybrid deep-convolutional neural network-based model (LungNet) was proposed by~\cite{ref1} with 53 total layers. The 19089 CT scan images were used in the study, and the data set was obtained from CancerImagingArchive.net. The LungNet model was compared to AlexNet, and the false positive was 0\% compared to AlexNet false positive which was 6.4\%.

\cite{mehta2021lung} proposed a hybrid approach, a 3D CNN, and a Random Forest. The authors use (LIDC-IDRI) dataset. The dataset has 1018 CT cases. Binary classification is used, benign and malignant. A K-Nearest Neighbors algorithm was used to identify nodules with features that are closest to benign or malignant. The analysis of their study shows that the Forest model that used only Biomarkers outperformed their novel hybrid model. In 2019,~\cite{xu2019deep} used CNN and RNN to help predict survival and measure other outcomes of patients’ diagnoses with NSCLS. The authors used their CNN model to predict survival using images prior to and post-radiation therapy. Using CNN and RNN models, in their study, they were able to track tumors and predict survival and prognosis at one and two years of overall survival.~\cite{kirienko2018convolutional} developed a CNN to classify lung cancer lesions.~\cite{Tekade2018LungCD} proposed a 3D multipath VGG-like network. The authors chose the VGG-like model because the model trains faster than other models. The architecture of the model VGG-16 has fully convolutional layers.

In 2018,~\cite{hosny2018deep} introduced a 3D CNN model. The study aims to employ a CNN model in predicting a two years survival of NSCLC patients. Their model is gauged in contrast to random forest models. However, the authors pointed out some limitations that the study has, such as the prognostics knowledge refined into the CNN model is based on past treatment options and plans and may not be suitable to predict prognostics for patients treated with the latest technology.~\cite{ardila2019end} developed a 3D CNN model to detect cancer regions in CT scans to predict cancer risk. Yunlang She et al.~\cite{she2020development} developed a deep learning model (DeepSurv) to predict survival among lung cancer patients. DeepSurv could be used as a useful tool for patients’ treatment recommendations. There are a few advantages for DeepSurv identified in the study: (I) DeepSurv is adaptive to variables such as real-world clinical factors. (II) Flexibility when dealing with complicated elements. (III) Learn and analyze censored features. (IV) Perform better in big data analysis. The study also indicated some limitations: (I) The excessive cost associated with deep learning models training and validating. (II) Interpreting the model’s prediction could be hard because the functionality of deep learning networks is like black boxes. (III) The study lacks external validation. In a study conducted by~\cite{song2017using}, three types of deep neural networks were proposed. Convolutional Neural Network (CNN), Deep Neural Network (DNN), and Stacked Autoencoder (SAE). The limitation of this study is pointed out by the authors in that the layers of the models are relatively small. In 2020,~\cite{yoo2020validation} aims to evaluate the performance of a deep learning model to detect lung cancer. In their study, ResNet34, a 34-layer CNN model is used to classify images. Table~\ref{compare_table} gives a comparison between the state-of-the-art CNN-based models for detecting lung cancer.

\section{Methodology}
In this paper, we proposed a deep convolutional neural network (CNN) model for early predicting and detecting different lung cancer stages. 

\subsection{Model architecture}
Our proposed model used several convolutional layers to perform the detection task. Table~\ref{modle_summary} shows the proposed model summary. The table shows the layers, the corresponding filters, the activation function, the output shape, and the number of parameters at each layer. The rectified linear unit (ReLU) has been used to add the nonlinearity~\cite{elsayed2018empirical}. The max-pooling layer was added to prevent overfitting~\cite{elsayed2018deep}. The SoftMax activation function~\cite{lecun2015deep} was used because classify three classes of lung cancer: benign, malignant, and normal. 

\begin{table*}[t]\centering
	\small
	\begin{tabular}{|lccclc|}
		\hline
		\textbf{Layer} & \textbf{Filters} & \textbf{Kernel} & \textbf{Activation} &\textbf{Output Shape}& \textbf{No. Param.} \\ 
		\hline 
		Input Conv2D        & 8 &  (3,3)& ReLU & (None,224,224,8) & 80\\
		\hline
		Max-pooling2D layer &  &  (2,2)&  & (None,112,112,8) & 0 \\
		\hline
		Conv2D &  16 &  (3,3)& ReLU & (None,110,110,16) &1168 \\
		\hline
		(Max-pooling2D layer &  &  (2,2)&  & (None,55,55,16) & 0 \\
		\hline
		Conv2D &  32 &  (3,3)& ReLU & (None,53,53,32) &4640 \\ 
		\hline
		Max-pooling2D &  &  (2,2)&  & (None,26,26,32) & 0 \\
		\hline
		Conv2D &  64 &  (3,3)& ReLU & (None,24,24,64) &18694 \\
		\hline
		Max-pooling2D&  &  (2,2)&  & (None,12,12,64) & 0 \\
		\hline
		Flatten &   &  &  & (None,9216) &0\\
		\hline
		Dense &   &  &  & (None,24) &221208\\
		\hline
		Dense &   &  & SoftMax & (None,3) &75\\

		\hline
		Total parameters &   &  &  & & 245667 \\ 
		\hline 
		Trainable parameters &   &  &  &  & 245667 \\ 
		\hline 
		
		Non-trainable parameters  &   &  &  & & 0 \\ 
		\hline 
	\end{tabular}
	\label{modle_summary}	
		\caption{The proposed CNN-based model for early prediction and detection of lung cancer stages summary.}
\end{table*}

\subsection{Dataset}
We used the dataset from IQ-OTH/NCCD-Lung Cancer Dataset in Kaggle~\cite{kareemiq}. The dataset is divided into three classes, benign, which contains 15 cases, malignant, which contains 40 cases; and normal, which contains 55 cases. The size of the dataset is 219 MB and contains 1097 CT scan images. Benign cases contain 120 images, malignant cases contain 561 images, and normal cases contain 416 images. We performed data preprocessing for the dataset images prior to using them for the model training stage.
\subsection{Data Preprocessing}
The dataset we used (IQ-OTH/NCCD-Lung Cancer)is real-world data and contains some quality issues, missing values, inconsistency, noise, and incompatible formats. Therefore, the data preprocessing stage is crucial to help eliminate inconsistencies, remove duplicates, and normalize the data.

\subsubsection{Image Resizing:} All the images in the dataset were resized to 224$\times$224. Prior to resizing, the benign class had 120 images with an image size of 512$\times$512, the malignant class had one image with a size of 404$\times$511, 501 images with an image size of 512$\times$512, 31 images with an image size of 512$\times$62, 28 images with an image size of 512$\times$801, and for the images in the normal class, one image with a size of 331$\times$506, 415 images with an image size of 512$\times$512.

\begin{figure}[t] 
	\centering
	\includegraphics[width=8cm, height=5cm]{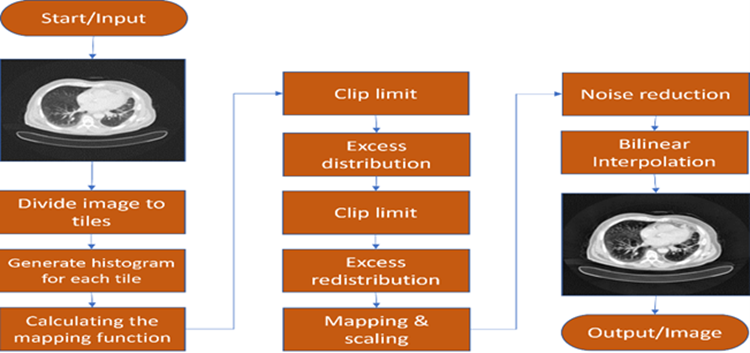}
	\caption{A flowchart of the image enhancement process using CLAHE.}
	\label{CLAHE}
\end{figure}

\begin{figure}[t] 
	\centering
	\includegraphics[width=8cm, height=3cm]{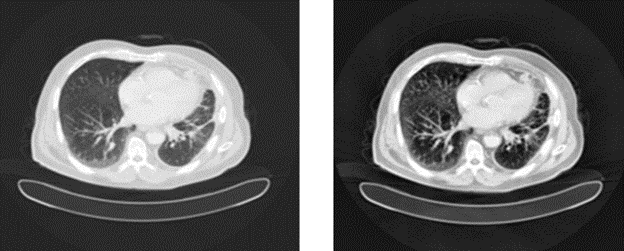}
	
	\caption{The input image before enhancement (on the left), and the output enhanced image (on the right).}
	\label{fig:my_label4}
\end{figure}

\subsubsection{Image Enhancement}
Contrast Limited Adaptive Histogram Equalization (CLAHE) is applied to enhance the contrast of the dataset images. CLAHE is a variant of adaptive histogram equalization in which the contrast amplification is limited, to reduce this problem of noise amplification. In CLAHE, the contrast amplification in the vicinity of a given pixel value is given by the slope of the transformation function.~\cite{clahe} CLAHE is a block-based image processing. CLAHE Algorithm works on a small area of the image called tiles, the contrast procedure is applied to each tile. After enhancing the contrast in each tile, they are combined by using a resampling technique called bilinear interpolation. Bilinear interpolation is used to prevent induced borderlines around the tiles. Figure~\ref{CLAHE} is a flow chart of the image enhancement process using CLAHE. There are five steps to enhance images using CLAHE:

\begin{enumerate}
	\item Each image is divided into square tiles as a two-factor vector of a positive integer. M represents rows, columns are represented by N, and T represents the total number of tiles. We used the tile default tile size in OpenCV 8$\times$8.
	\begin{equation}
		T = M*N
	\end{equation}
	\item The mapping function of local histogram is calculated by: 
	\begin{equation}
		f(x_j) = (G - 1) \sum_{i=0}^{j} \frac{t_i}{T}
	\end{equation}
	where(G) represents total possible gray levels and T represents the total of pixels in an image. The cumulative number of gray pixels between zero and $\mathrm{X_j}$ is represented by: $\sum_{i=0}^{j} \frac{t_i}{T}$
	\item We use the clipping point of CLAHE. Clip limit, also called contrast-enhanced limit, is used to normalize an image and prevent over-concentration in similar areas of the image. We can use a clip limit value between zero and one. The higher the value, the more contrast. The clip limit default value is 0.01. The clip limit determines how many pixels are allowed in a histogram bin. The default number of histogram bins is 256. It is a positive integer and is used to develop a contrast-enhancing transformation.
	\item We apply the histogram equalization to each region. 
	\item To reduce the noise in the image. Median filters are used to remove the noise and preserve the image edges. This is essential because the noise could lead to a false positive. Figure~\ref{fig:my_label4} is an example of an input image before enhancement and the output enhanced image using CLAHE. Any extra noise must be removed to prevent any false positives due to noise presence.
\end{enumerate}

\subsection{Data augmentation}
Data augmentation has been performed on the dataset images to increase the dataset size and increase the problem complexity. The following augmentations have been performed:
\begin{itemize}
	\item	Flip-top-bottom is 40\%. 
	\item	Flip-left-right is 30\%. 
	\item	Random brightness with a probability of 30\%, minimumf actor of 30\%, and maximum factor of 1.2. 
	\item	Zoom random is 20\%. 
	\item	Also, we used horizontal flip. 
	\item	The rotation range is 40.
	\item	Shear range is 20\%. 
	\item	Rescaling by 1/255. 
	\item	Height shift range is 20\%.
\end{itemize}
Then, data is split into three sets 70\% training set, 15\% validation set, and 15\% testing set.

\begin{figure}[htp] 
	\centering
	\includegraphics[width=6cm, height=4cm]{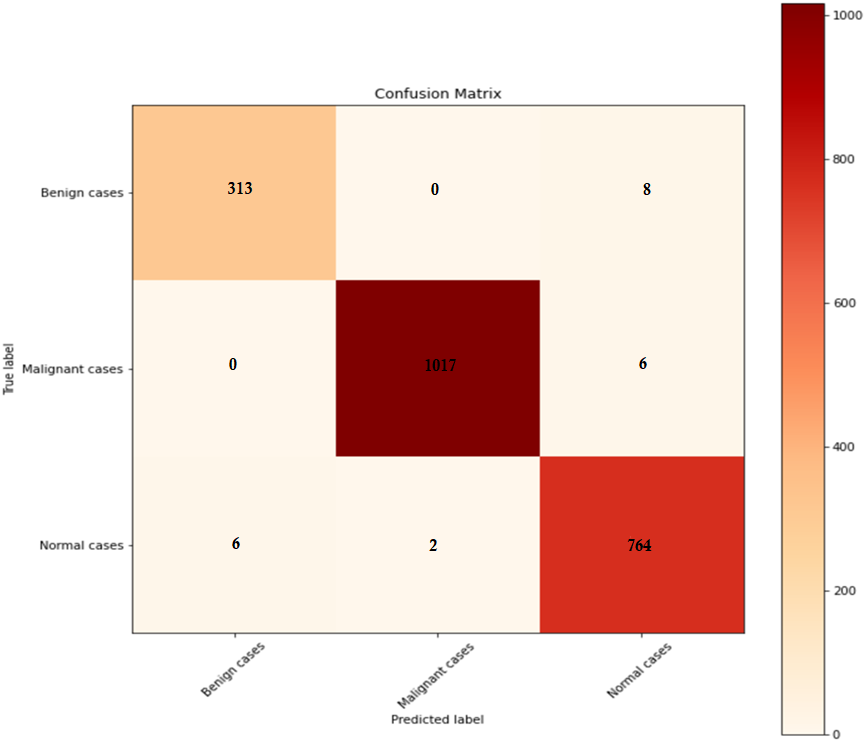}
	\caption{The confusion matrix for the proposed model to detect the lung cancer.}
	\label{fig:my_label2}
\end{figure}

\begin{table*}[t]
	\centering
	\begin{tabular}{|llllll|}
		\hline
		\textbf{Class }& \textbf{Precision} & \textbf{Sensitivity)}& \textbf{Specificity} &\textbf{F1-score}& \textbf{Support} \\ 
		\hline 
		Benign (0) &	98\%&	98\%&	99\%&	98\%&	321\\
		\hline
		Malignant (1) &	1.00&	99\%&	99\%&	1.00&	1023 \\
		\hline
		Normal (2)&	98\%&	99\%&	98\%&	99\%&	772 \\
		
		\hline
		Accuracy& & &	 	 	 	&99\%&	2116 \\
		\hline
		Loss& & &	 	 	 	&0.17&	2116 \\ 
		\hline
		Macro Average&99\% &99\% &	99\% 	 	 	&99\%&	2116 \\
		\hline
		Weighted Average&99\% &99\% &99\%	 	 	 	&99\%&	2116 \\
		\hline
		\end{tabular}	
	\caption  {The proposed model empirical results.}
	\label{results_table}
\end{table*}
\section{Results and Discussion}
Our model was built in Python 3 using TensorFlow-Keras libraries. We have run the model on Google Colab Pro with Tesla P100-PCIE-16GB, and 32 GB RAM. To measure the performance of our CNN model, we used a variety of metrics.

In our experiment, the original dataset consists of one thousand ninety-seven (1097) CT scan images. We have increased the dataset to eight thousand 8461 images using augmentation methods. Six thousand three hundred forty-five images were used for training, and 2116 images were used for validation. The training images were divided into three classes; benign cases consisted of 961 images, malignant cases consisted of 3067 images, and normal cases consisted of 2317 images. Also, the validation images were divided into three classes; benign cases consisted of 321 images, malignant cases consisted of 1023 images, and normal cases consisted of 772 images. 

Several metrics have been used to measure our model performance. Figure~\ref{fig:my_label2} shows the confusion matrix diagram for the three classes where the True Positive (TP) indicates the number of times where the actual value is yes (true), and the model predicted yes. False Positive (FP) indicates the number of times where the actual value is yes (true), and the model predicted yes. This is known as a Type-I error. True Negative (TN) indicates the number of times the actual value is no, and the model predicted no. False Negative (FN) indicates the number of times the actual value is yes, and the model predicted no. This is known as a Type-II error.
\begin{figure}[t] 
	\centering
	\includegraphics[width=8cm, height=3cm]{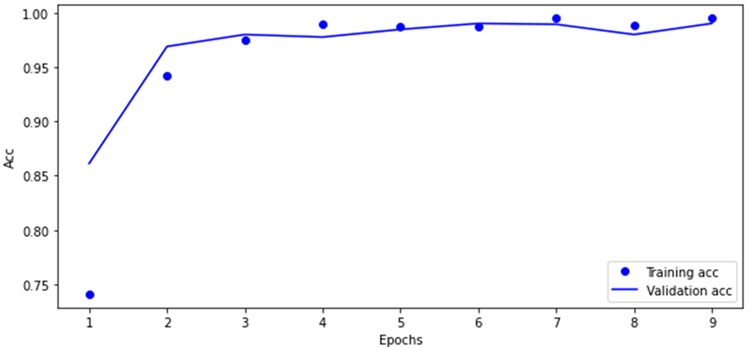}
	
	\caption{The proposed model training versus validation accuracies.}
	\label{accuracy_diagram}
\end{figure}

\begin{figure}[t] 
	\centering
	\includegraphics[width=8cm, height=3cm]{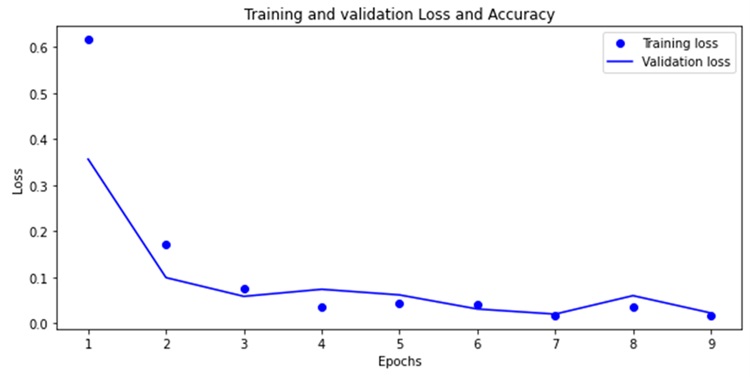}
	
	\caption{The proposed model training versus validation loss.}
	\label{loss_diagram}
\end{figure}

The accuracy of the proposed model has been calculated by: 

	\begin{equation}
	Accuracy  =  \frac{TP + TN}{TP + FP + TN + FN}
\end{equation}

The precision value that reveals how many of the predicted positive values are actually true is calculated by: 
	\begin{equation}
	Precision  =  \frac{TP}{TP + FP}
\end{equation}
The recall (sensitivity) that represents the actual output that is predicted correctly is calculated by:

\begin{equation}
	Recall(Sensitivity)  =  \frac{TP}{TP + FN}
\end{equation}

The specificity that shows the true negative of the model by applying the following formula: 
	
	\begin{equation}
		Specificity= \frac{TN}{TN + FP}
	\end{equation}
Combining the recall and precision of the model's classification into a single metric. To maintain the balance between recall and precision, we use the harmonic mean:
	
	\begin{equation}
		F1-Score  =  2* \frac{Recall * Precision}{Recall + Precision}
	\end{equation}

Precision for benign, malignant, and normal cases are 98\%, 1.00, and 98\%, respectively. The results of sensitivity (recall) for benign, malignant, and normal cases are 98\%, 99\%, and 99\%, respectively. Specificity for benign, malignant, and normal cases are 99.55\%, 99.45\%, and 98.65\%, respectively. F1-score results for benign, malignant, and normal cases are 0.98, 1.00, and 99\%, respectively. Measuring The macro average and the weighted average for Precision, recall, and F1-score, the output was the same 99\%. The model's accuracy is 99.45\%, and the loss is 1.75\%. Figure~\ref{accuracy_diagram} and Figure~\ref{loss_diagram} show the proposed CNN model's accuracy and loss for the training versus validation diagrams over the ten epochs. Table~\ref{results_table} shows the proposed model oveall statistical results.

In our experiment, the original dataset consists of one thousand ninety-seven (1097) CT scan images. We have increased the dataset to eight thousand four hundred sixty-one (8461) images using augmentation methods. Six thousand three hundred forty-five (6345) images were used for training, and two thousand one hundred sixteen (2116) images were used for validation. The training images were divided into three classes, benign cases consisted of nine hundred sixty-one (961) images, malignant cases consisted of three thousand sixty-seven (3067) images, and normal cases consisted of two thousand three hundred seventeen (2317) images. Also, the validation images were divided into three classes, benign cases consisted of three hundred twenty-one (321) images, malignant cases consisted of one thousand twenty-three (1023) images, and normal cases consisted of seven hundred seventy-two (772) images. The performance of the CNN model was assessed using the confusion matrix. Table 4.4 shows the confusion matrix report. Precision for benign, malignant, and normal cases are 98\%, 1.00, and 98\%, respectively. The results of sensitivity (recall) for benign, malignant, and normal cases are 98\%, 99\%, and 99\%, respectively. Specificity for benign, malignant, and normal cases are 99.55\%, 99.45\%, and 98.65\%, respectively. F1-score results for benign, malignant, and normal cases are 0.98, 1.00, and 99\%, respectively. Measuring The macro average and the weighted average for Precision, recall, and f1-score the output was the same 99\%. The model's accuracy is 99.45\%, and the loss is 1.75\%. Figure 4.5 shows the proposed CNN model’s accuracy and loss. 

We have compared our results with three different states of the art models. Table~\ref{compare_results} shows a comparison between our proposed model and some of the state-of-
the-art models used the same dataset (IQ-OTH/NCCD).
\begin{table*}[t]
	\centering
	\begin{tabular}{|llllllll|}
		\hline
		\textbf{Author}&	\textbf{Year}&\textbf{Dataset}&\textbf{Model}&	\textbf{Epoch}&\textbf{Sensitivity}	&\textbf{Specificity}&\textbf{Accuracy} \\
		\hline
\cite{kareem2021evaluation}&	2020&	IQ-OTH/NCCD&	SVW& N/A&	97.14\%&	97.5\%&	89.88\% \\
		\hline
		\cite{al2021transfer}&	2021&	IQ-OTH/NCCD&	GoogleNet&N/A& 95.08\%&	93.7\%&	94.38\% \\
		\hline
\cite{al2020diagnosis} &	2020&	IQ-OTH/NCCD	&AlexNet&100&	95.71\%	&95\%&	93.45\% \\
		
		
	
		\hline
		Proposed model& & IQ-OTH/NCCD& CNN& 10&	99\%&	99.21\%	& 99.45\% \\
		\hline 
		
	\end{tabular}
	\label{compare_results}
	\caption  {A comparison between the proposed model and convolutional-based state-of-the art models over the IQ-OTH/NCCD-Lung Cancer Dataset.}
\end{table*}


\section{Conclusion}
This paper proposed a CNN-based model for early prediction and detection of lung cancer from CT scan imaging. The model detected benign, malignant, and normal cases. Detecting lung cancer at an early stage is crucial, and this will help medical providers to start their treatment plan, which will increase the survival rate. One of the objectives that our model achieved is the reduction of false positives. Moreover, the proposed model has achieved a high accuracy rate of 99.45\%. Lung cancer is one of the most difficult cancers diagnosed in an early stage. With the existing methods that radiologists use, the proposed model in this thesis can be a particularly useful tool to support their decisions.

\bibliography{referenceslungsPaper}
\bibliographystyle{flairs}


\end{document}